\documentclass[a4paper,jou,natbib]{apa6}
\usepackage[english]{babel}
\usepackage[draft]{hyperref}

\hypersetup{
    colorlinks,
    linkcolor={black},
    citecolor={black},
    urlcolor={black}
}
\usepackage{graphicx}

\usepackage{natbib}

\newcommand{\refsubfig}[2]{\hyperref[#1]{\autoref{#1}#2}}

\makeatletter
\renewcommand{\mspart}{{\ifapamodeman{\clearpage}{}}\@startsection {section}{1}{\z@}%
    {\b@level@one@skip}{\e@level@one@skip}%
    {\centering\normalfont\normalsize\bfseries}}
\makeatother

\title{Gerrymandering and computational redistricting}

\threeauthors{Olivia Guest}{Frank J. Kanayet}{Bradley C. Love}
\threeaffiliations{Department of Experimental Psychology,\\ University College
London, UK}{Department of Psychology,\\ Stanford, USA}{Department of
Experimental Psychology,\\ University College London, UK\\ \& The Alan Turing
Institute,\\ London, UK}
\leftheader{Guest, Kanayet, and Love}
\abstract{
Partisan gerrymandering poses a threat to democracy. Moreover, the complexity
of the districting task may exceed human capacities. One potential solution is
using computational models to automate the districting process by optimizing
objective and open criteria, such as how spatially compact districts are. We
formulated one such model that minimized pairwise distance between voters
within a district. Using US Census Bureau data, we confirmed our prediction
that the difference in compactness between the computed and actual districts
would be greatest for states that are large and therefore difficult for humans
to properly district given their limited capacities. The computed solutions
highlighted differences in how humans and machines solve this task with
machine solutions more fully optimized and displaying emergent properties not
evident in human solutions. These results suggest a division of labour in which
humans debate and formulate districting criteria whereas machines optimize the
criteria to draw the district boundaries. We discuss how criteria can be
expanded beyond notions of compactness to include other factors, such as
respecting municipal boundaries, historic communities, relevant legislation,
etc.}

\authornote{The authors would like to thank Christiane Ahlheim, Kurt Braunlich,
Adam Hornsby, Vladimir Sloutsky, and Sebastian Bobadilla Suarez for their
feedback on the manuscript. B.C.L was supported by Leverhulme Trust grant
RPG-2014-075 and Wellcome Trust Senior Investigator Award WT106931MA. O.G.
wrote the manuscript, created the figures, programmed the implementation, and
developed the theory and concept; F.J.K. programmed the prototype, and tested
the original concept; B.C.L. wrote the manuscript, and developed the theory and
concept. The authors declare that they have no competing financial interests.
The data and code are archived at the following address:
\href{https://osf.io/5fepu/}{https://osf.io/5fepu/}. Correspondence should be
addressed to O.G. (email: o.guest@ucl.ac.uk).}

\begin{document}
\maketitle

One of the greatest threats to democracy, particularly in the USA, is
gerrymandering. Gerrymandering is the practice of (re)drawing electoral district
boundaries to advance the interests of the controlling political faction.
The
term is a portmanteau, coined in 1812 when people noticed
that a district --- approved by the then governor of Massachusetts, Elbridge
Gerry --- resembled a salamander \citep{Martis2008}.

\begin{figure}[ht!]
  \centering
   \includegraphics[width=50mm]{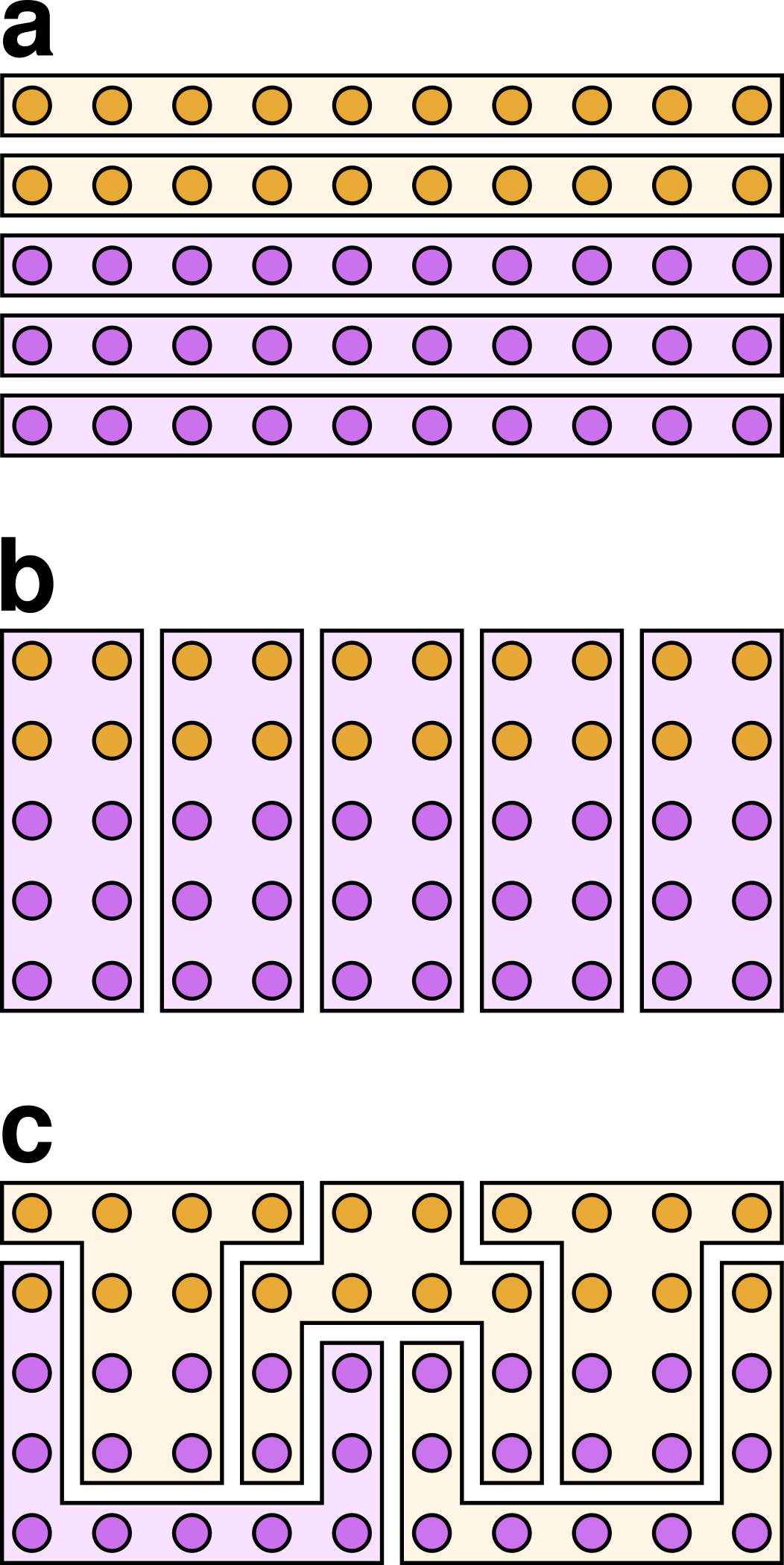}

   \vspace{5mm}

  \caption{An illustrative example of three redistricting plans. The 50 voters
  (circles) are grouped into 5 districts (polygons) with the background colour
  denoting the winning party. The purple party (60\% of voters) secures 60\%,
  100\%, and 40\% of the seats under the three plans, respectively. \textbf{a})
  Compact, fair: the proportion of wins (60\%) by the purple party reflects its
  overall level of voter support. \textbf{b}) Compact, not fair: all five
  districts are won by the purple party because the orange vote has been
  cracked. \textbf{c}) Not compact, not fair: the purple party has been packed
  into two districts (its only wins) and cracked in the remaining districts. We
  recommend the video that motivated this figure \citep{taylor2016}.}
  \label{fig:gerrymandering}
\end{figure}

Gerrymandering leads to districts with unnecessarily visually complex shapes,
e.g., North Carolina (see \autoref{fig:north_carolina}c). Although there are
laws (both at the state and federal level) to safeguard the rights of citizens
(including minorities) during the redistricting process, in practice these laws
do little to reduce partisan gerrymandering \citep{issacharoff2002}. Worryingly,
gerrymandering is on the rise \citep{stephanopoulos2015} due to partisan actions
of both Republicans and Democrats \citep{bazelon2017}. In the 17 states where
Republicans controlled the redistricting process, they secured 72\% of the
available seats on only 52\% of the vote. Mirroring, in the 6 states where
Democrats controlled the districting process, they secured 71\% of the seats on
56\% of the vote.

The two main gerrymandering strategies are \emph{packing} and \emph{cracking}
\citep{altman2015}. Cracking dilutes people likely to vote for the opposition,
assigning them to as many districts as possible, see
\autoref{fig:gerrymandering}b. One cracking tactic is to dilute urban voting
blocs by having multiple districts from the countryside converge like the spokes
of a wheel at a city's fractured hub. In contrast, packing concentrates people
who will likely vote for the opposition within a small number of districts,
rendering their vote inconsequential in the remaining districts, see
\autoref{fig:gerrymandering}c.

One possible solution to partisan gerrymandering is to rely on computer
algorithms to impartially draw districts \citep{vickrey1961}. In theory, there
is no reason why such a solution could not be adopted in the USA. Indeed, many
states within Mexico use computer algorithms to district \citep{altman2014a,
garcia2015, gutierrezandrade2016, ponsich2017}. Moving to computational
redistricting would ``elevate the legislative redistricting debate from a battle
over line drawing to a discussion of representational goals'' \cite[p.
1381]{browdy1990}. In other words, the role for humans would be to decide and
formalise the criteria (e.g., people within a district should be close to one
another) and the computer's job would be to find the best solution without human
tinkering. Thus, the appeal of automation is twofold: \textit{a}) open-source
computer algorithms can be written to follow objective criteria absent
corrupting influences; and \textit{b}) computers are able to toil away
optimising the objective criteria in contrast to humans who have limited
cognitive capacity and time to devote. While the first point, namely that
purposeful gerrymandering occurs and leads to unfair solutions may be obvious,
the second point may be less so. However, from a psychological perspective, it
is clear that humans do not consider all logically possible solutions in
combinatoric problems (e.g., districting), but may instead rely on shortcuts and
general organisational principles \citep{Palmer1990}. Counterintuitively, some
of what we perceive as gerrymandering may simply reflect that humans are not
very good at the districting task.

In light of these observations, we tested the psychologically-motivated
prediction that differences in compactness between the computed and actual
districts will be greatest for states that are large and therefore difficult for
humans to properly district. To evaluate this prediction, we devised a novel
clustering algorithm to redistrict the USA's 435 congressional districts (more
populous states are allotted more seats). The algorithm maximises a notion of
compactness by minimising the average mean distance between people within the
same district, cf. \citep{hess1965, arnold2017, chen2013}

In accord with federal law, our novel algorithm includes an additional
constraint to create clusters (i.e., districts) of roughly the same cardinality
(i.e., population). We refer to our algorithm as weighted $k$-means because it
is based on $k$-means clustering \citep{arthur2007, lloyd1982}. Details are
provided in the Methods section and the open-source code is available to
reproduce the reported results at
\href{https://osf.io/5fepu/}{https://osf.io/5fepu/}.

\section*{Materials and Methods}
This section details how the US Census Bureau data was preprocessed, provides
details on the weighted $k$-means model.

\subsection*{Census Data}
US Census Bureau data were used to perform the district clusterings reported in
the main text. For clustering, we used the smallest available geographic unit,
known as a census block. The US Census Bureau collects data for just over 11
million census blocks of which almost 5 million have a population of 0. The last
decennial census occurred in 2010. However, as recently as 2015, the US Census
Bureau conducted the ACS (American Community Survey), which is a survey at one
level above the block level, which is referred to as a block group. Using these
2015 counts, we estimated the population of each census block in 2015 by
calculating its population share of its block group in 2010 and, assuming these
proportions had not changed, updated the block populations based on the 2015
ACS. Notice that our population estimates for census blocks in 2015 is not
constrained to be an integer.

Census blocks in urban areas tend to be geographically smaller but more
populated. Based on our estimates combining the 2010 and 2015 data, the mean
population of a census block is $29.49$ with a median of $3.41$ people. The mean
area of a census block $1.11$ km$^2$ with a median of $0.04$ km$^2$.

\subsection*{Initialisation}
The manner in which clusters are initialised will affect the quality of the
final solution because our algorithm, like $k$-means which it generalises, moves
toward a local optimum. We initialise the centroids using the procedure from
$k$-means++ \citep{arthur2007}.

\subsection*{Weighted $k$-means Algorithm}
Weighted $k$-means generalises $k$-means by preferring clusters of roughly the
same cardinality (i.e., number of members) with the strength of this preference
determined by a parameter value. Like $k$-means, in each iteration, items are
assigned to the nearest cluster and at the end of iteration the position of the
cluster (i.e., centroid) is updated to reflect its members' positions. After a
number of iterations, the algorithm converges to a local optimum. Weighted
$k$-means differs from $k$-means by penalising clusters with more members such
that distances to these clusters are multiplied by a scaling factor reflecting
the cluster's cardinality. The weight for cluster $i$ is
\begin{eqnarray}
  \label{eq:w}
  w_i = \frac{|C_i|^{\alpha}} {\displaystyle\sum_{j=1}^{K}{|C_j|^{\alpha}}} ,
\end{eqnarray}
where $|C_i|$ is the
cardinality of cluster $i$, $K$ is the number of clusters, and $\alpha$ is a
parameter that determines how much to penalise clusters with a disproportionate
number of members.

To stabilise solutions across iterations and prevent oscillations, the scaling
factor ${s}_{i, t}$ for cluster $i$ at time $t$ (i.e., iteration $t$) is
calculated as a weighted combination of the previous scaling factor ${s}_{i,
t-1}$ and $w_i$
\begin{eqnarray}
  \label{eq:s}
  {s}_{i, t}= \beta {s}_{i, t-1} + (1-\beta) w_i ,
\end{eqnarray}
where $\beta$ is a control parameter in the range
$[0, 1)$. In the first iteration, each ${s}_{i, 0}$ is initialised to
$\frac{1}{K}$ where $K$ is the number of clusters.

The scaled distance of point $x$ to the cluster $i$ is
\begin{eqnarray}
  \label{eq:d_x}
  d_{s}(x, \mu_i) = s_{i,t} \times d(x, \mu_i) ,
\end{eqnarray}
where $\mu_i$ is the position of cluster $i$ and $d$ is the
distance metric, which in this contribution is great-circle distance (also known
as orthodromic or geodesic distance, estimated using the haversine formula),
which respects the curvature of the Earth.

Finally, $\textup{argmin}_i d_{\textup{s}}(x, \mu_i)$ is used to find the
nearest weighted cluster $i$ from point $x$, to which $x$ will be assigned.
Notice that this algorithm is identical to $k$-means when $\alpha$ is $0$. As
$\alpha$ increases, the constraint of equal cardinality becomes firmer.

\subsection*{Parameter Fitting}
Solutions are only considered that converge and for which the cardinalities of
the clusters are in line with that of actual congressional districts.  In
principle, one could use any parameter search procedure to find $\alpha$ and
$\beta$ that minimised the measure we report, which is the pairwise distance of
voters within a district (i.e., cluster). For example, one could use grid search
to consider all possible combinations (at some granularity) of $\alpha$ and
$\beta$.

However, given available computing resources, we adopted a more efficient
procedure informed by our understanding of the algorithm's behaviour (i.e.,
smaller $\alpha$ values lead to tighter clusterings). The parameter search
procedure began with $\alpha$ set to 0 and increased $\alpha$ until an
acceptable solution was found. At each level of $\alpha$, $\beta$ was set to
$.5$ and increased by $.1$ after a simulation failure until $\beta$ exceeded its
range. At that point, $\alpha$ was increased by $.1$ and the process was
repeated with $\beta$ set to $.5$. This procedure terminated when an acceptable
solution was found. At that point, a finer-grained optimisation was performed,
which considered $\alpha$ values up to $.1$ lower than first acceptable value
found.

\section*{Results}
\begin{figure*}[ht!]
  \centering
  \includegraphics[width=120mm]{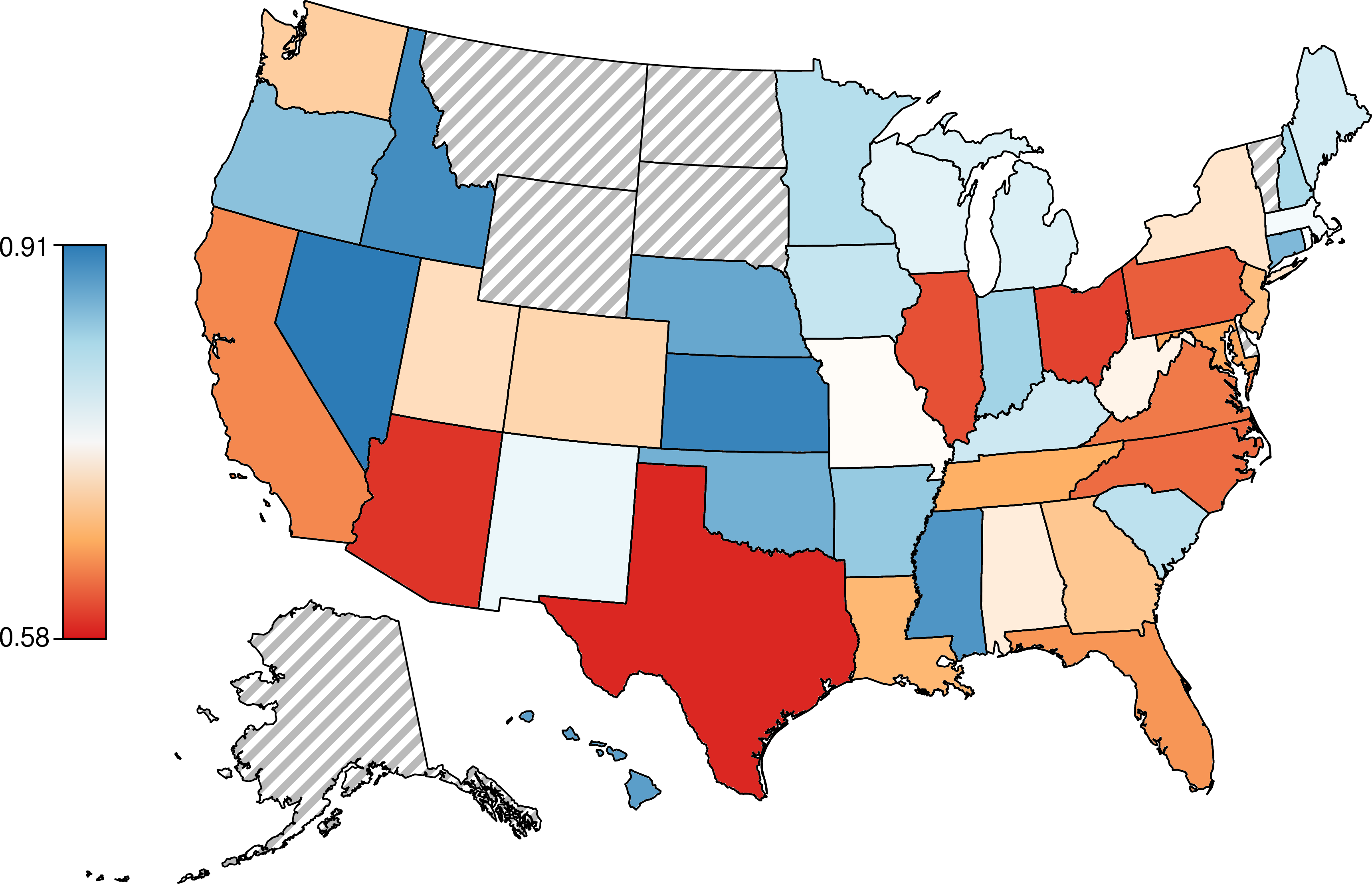}

  \vspace{5mm}

  \caption{Map of the USA showing how much more compact each state's districts
  would be under computational districting. Red-coloured states would improve
  the most after using our algorithm to form districts that are compact by
  minimising the pairwise distances between people within a district.
  Blue-coloured states would improve the least from computational redistricting,
  though still show an improvement in within district pairwise distances. States
  with grey hatching, e.g, Alaska (bottom left), have only one district.}
  \label{fig:whole_usa}
\end{figure*}

Our clustering algorithm created improved maps for every state, see
\autoref{fig:whole_usa}. Please visit
\href{http://redistrict.science}{http://redistrict.science} to compare the
actual and automated districting plans for any address in the USA. We define the
improvement for each state as the ratio of pairwise distances within districts
between our solution and the actual districts. This metric favours districts in
which voters are tightly clustered spatially, leading to a mean improvement
across states of  $0.796$ (i.e,. about 20\%) with standard deviation $0.0858$.
To test our main prediction, a regression model was fit to the state improvement
scores with number of districts, and square of number of districts serving as
predictors, $R^2 = 0.550$, $F(2, 40) = 24.47,\, p \approx 0$. Both predictors in
the fit, $-\, 0.0149(\textup{number of districts})\, +\, 0.0002(\textup{number
of districts})^2 +\, 0.9027$, were statistically significant, $t(40) = -5.654,\,
p \approx .0$ and  $t(40) = 3.879,\, p \approx 0$, respectively. Consistent with
our prediction, these results suggest that the cognitive demands of drawing
districts for larger states may tax human capacities. Thus, some of the
unfairness in current solutions may be unintentional, as opposed to wholly
attributable to deliberate gerrymandering for political gain.

The regression also included a quadratic term for the number of districts that
confirmed the intuition that the complexity of the task should not scale
linearly with the number of districts because the clustering is spatial and
local interactions dominate. For instance, there are natural groupings and
locality within big states, e.g., what is drawn for Los Angeles is unlikely to
strongly affect what is drawn in San Francisco. People likely segment maps
hierarchically into regional groupings to reduce processing demands, as they do
in other map reasoning tasks \citep{graham2000}, which may explain fallacious
conclusions like that Reno lies east of Los Angeles and that Atlanta is east of
Detroit \citep{myers2002}. Overall, the district size results indicate that
states with fewer districts are easier to draw properly, which suggests that
state size may be another cause of ``accidental gerrymandering''
\citep{chen2013}. The residuals from this regression model can be interpreted as
how gerrymandered each state is, adjusting for population. This analysis
suggests that Arizona is the most gerrymandered state (see
\autoref{tbl:residuals} for the complete ranking).

\begin{table}[!ht]
  \centering
  \caption{
  {\bf States sorted by their residuals from the regression model
  described in the main text. A state's residual can be interpreted as how
  gerrymandered the state is after taking into account the number of districts,
  with negative residuals indicating greater gerrymandering. Of course, there
  could be other important covariates in addition to population size.}}
  \begin{tabular}{crcrcr}
        \hline
        State & Residuals & State & Residuals & State & Residuals \\ \hline
        AZ    & -0.1482 & ME   & -0.0329 & NE   & 0.0267  \\
        MD    & -0.0820 & NM   & -0.0220 & OR   & 0.0399  \\
        LA    & -0.0792 & NH   & -0.0158 & SC   & 0.0401  \\
        OH    & -0.0747 & WA   & -0.0122 & WI   & 0.0419  \\
        VA    & -0.0747 & NJ   & -0.0043 & CT   & 0.0426  \\
        UT    & -0.0632 & CA   & -0.0036 & MA   & 0.0437  \\
        TX    & -0.0623 & IA   & -0.0022 & MS   & 0.0458  \\
        NC    & -0.0551 & AL   &  0.0027 & OK   & 0.0527  \\
        IL    & -0.0538 & HI   &  0.0137 & MN   & 0.0563  \\
        TN    & -0.0503 & GA   &  0.0181 & FL   & 0.0568  \\
        PA    & -0.0475 & KY   &  0.0203 & KS   & 0.0603  \\
        WV    & -0.0466 & ID   &  0.0217 & NV   & 0.0628  \\
        RI    & -0.0458 & MO   &  0.0239 & IN   & 0.0813  \\
        CO    & -0.0386 & AR   &  0.0244 & MI   & 0.1047  \\
              &         &      &         & NY   & 0.1345  \\ \hline
  \end{tabular}
  \label{tbl:residuals}
\end{table}

\begin{figure*}[ht!]
  \centering
  \includegraphics[width=120mm] {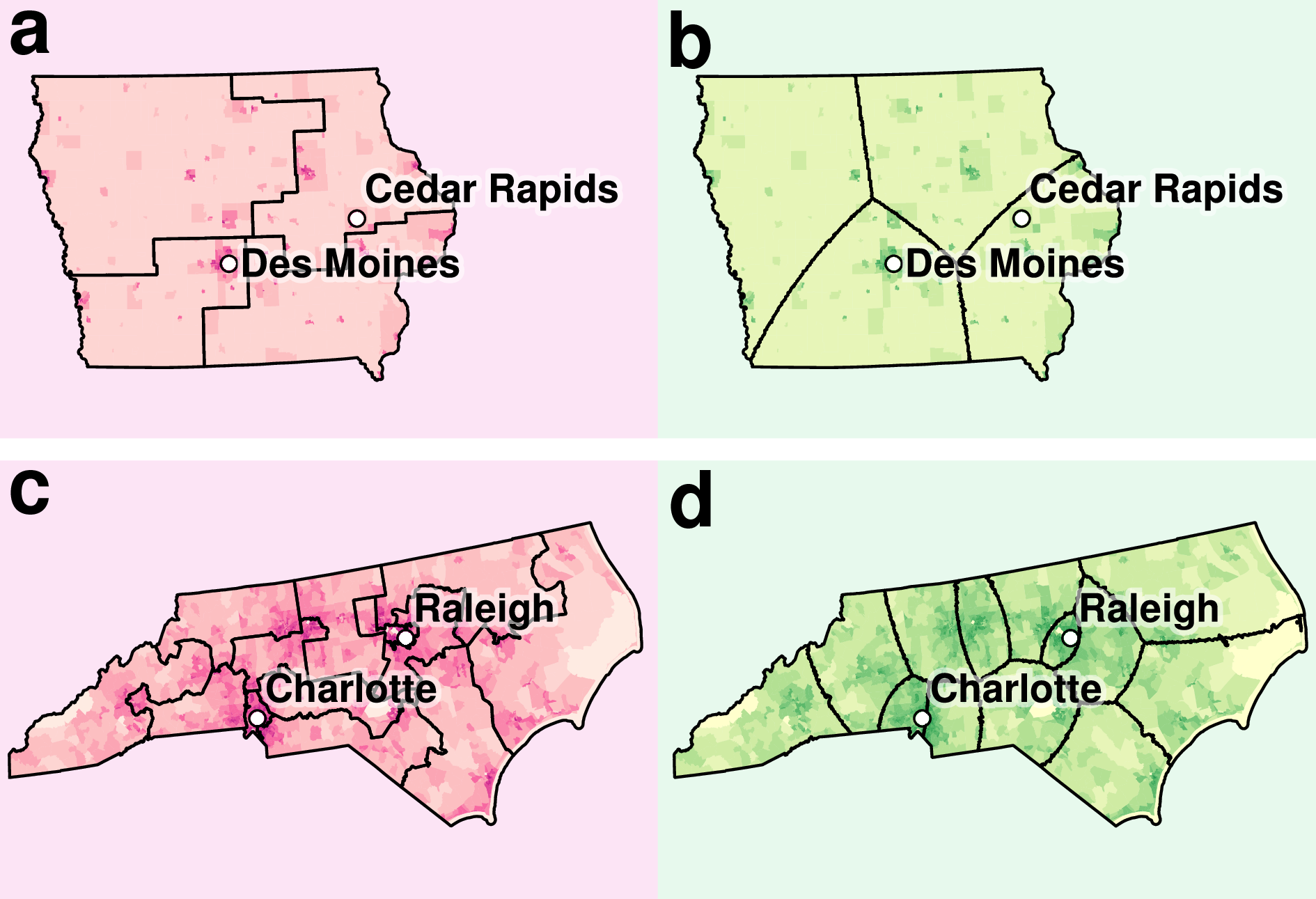}

  \vspace{5mm}

  \caption{Actual and computed district maps for Iowa (\textbf{a}, \textbf{b})
  and North Carolina (\textbf{c}, \textbf{d}). Computed solutions are shown in
  green to the right of the actual congressional districts. Darker areas on the
  map (census tracts) are more densely populated.}
  \label{fig:iowa}
  \label{fig:north_carolina}
\end{figure*}

Let us turn to some specific examples for redistricting solutions (for an
interactive map, visit
\href{http://redistrict.science}{http://redistrict.science}). For Iowa, which
uses a neutral commission to draw district boundaries \citep{levitt2011}, our
automated solution uses fewer segments (\autoref{fig:iowa}b) than the more
complex actual solution (\autoref{fig:iowa}a). In the case of North Carolina,
where maps are drawn through a partisan process, improvements are also evident
(\autoref{fig:north_carolina}c, d).

\begin{figure*}[ht!]
  \centering
  \includegraphics[width=120mm] {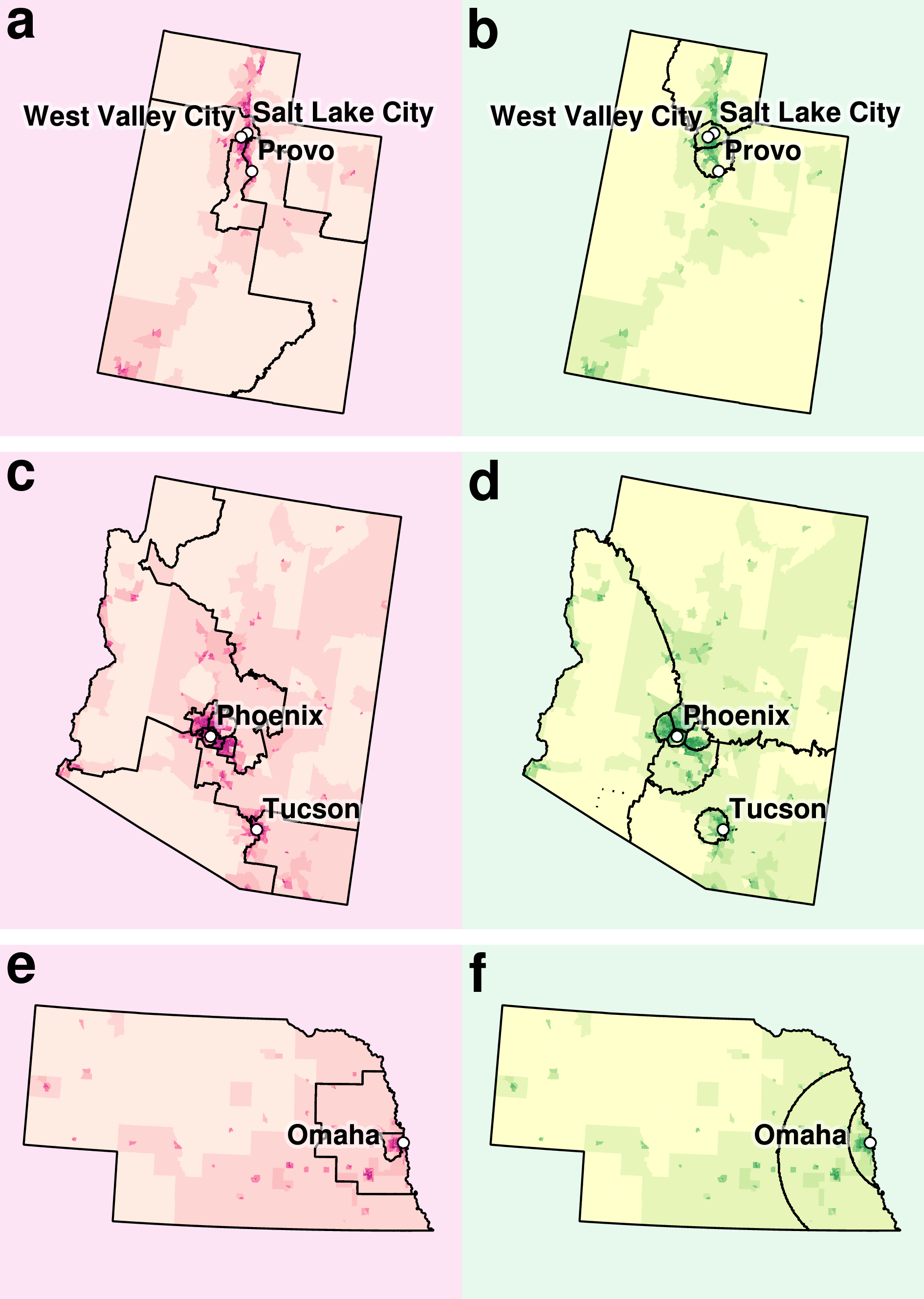}

  \vspace{5mm}

  \caption{Actual and computed district maps for Utah (\textbf{a}, \textbf{b}),
  Arizona (\textbf{c}, \textbf{d}), and Nebraska (\textbf{e}, \textbf{f}).
  Computed solutions are shown in green to the right of the actual congressional
  districts. Darker areas on the map (census tracts) are more densely
  populated.}
  \label{fig:utah}
  \label{fig:arizona}
  \label{fig:nebraska}
\end{figure*}

Notwithstanding Utah's ``long tradition of requiring that districts be [...]
reasonably compact'' \citep{christensen2001}, the densely populated northern
conurbation of Provo, Salt Lake City, and West Valley City, is cracked, diluting
the urban vote by recruiting parts of the countryside, reaching to the southern
border of the state (\autoref{fig:utah}a). In the computed solution, the urban
area of West Valley and Salt Lake City is assigned to a single urban district,
as is Provo and its surrounding conurbation (\autoref{fig:utah}b).

The automated districting of Arizona showcases an emergent property of our
algorithm that human-drawn maps have not displayed, namely that districts can be
embedded within one another, such as a small densely populated urban district
encircled by a large sparsely populated rural district (i.e., shaped like a
doughnut). Rather than crack Tuscon across 3 districts (\autoref{fig:arizona}c),
the algorithm settled on a doughnut structure (\autoref{fig:arizona}d).

An interesting case of convergence between human and algorithm is the case of
Nebraska (\autoref{fig:nebraska}e). Our algorithm followed in the footsteps of
those who districted Nebraska (\autoref{fig:nebraska}f), capturing the same
transition from fully urban (east) to fully rural (west). However, the smooth
radiating boundaries surrounding the capital, Omaha, are more compact
(optimised) in the automated solution.

One question is which solution potential voters prefer. In the first 3 days
\href{http://redistrict.science}{http://redistrict.science} was live, 367
self-identified US citizens indicated whether they preferred our algorithmic
solution or the actual districting for their state. The vast majority of
respondents preferred the computational solution (90.7\% overall; 91.1\% when IP
address location and state matched) with the pattern holding across states.

\section*{Discussion}
In summary, we applied our novel weighted $k$-means algorithm to US Census
Bureau data to redistrict the USA's 435 congressional districts and compared the
computed solutions to actual districts. The results confirmed our prediction
that larger states would tend to show greater improvement, suggesting that the
complexity of the districting task may overwhelm humans' ability to find optimal
solutions. One startling conclusion is that some of what we view as purposeful
gerrymandering may reflect human cognitive limitations. At this juncture, this
conclusion is more a provocative conjecture than an established finding. Further
work is needed to evaluate how human cognitive biases and limitations may
contribute to gerrymandering.

In light of our results, we advocate a division of labour between human and
machine. Stakeholders should openly debate and justify the districting criteria.
Once the criteria are determined by humans, it should be left to the computers
to draw the lines given humans' cognitive limitations and potential partisan
bias. We offer one of many potential solutions. The computer code, like ours
used in these simulations, should be open-source (to allow for replication and
scrutiny) and straightforward to provide confidence in its operation.

Political, ethical, scholarly, and legal debate should play a central role in
determining the optimisation criteria. For example, instead of choosing the mean
pairwise distance between constituents, we could have used travel time to
capture the effects of geographical barriers, such as rivers. Even a measure as
simple as travel time raises a number of ideological considerations that should
be debated, such as the mode of transportation (e.g., public, on foot, or by
automobile) to adopt. Other factors could be included in the criteria, such as
respecting municipal boundaries, historic communities, the racial composition of
districts, partisan affiliation, etc. For our demonstration, we chose perhaps
the simplest reasonable criteria, but in application the choice of criteria
would ideally involve other factors after lengthy debate involving a number of
stakeholders. These debates should elevate democratic discourse by focusing
minds on principles and values, as opposed to how to draw maps for partisan
advantage.

Although we focused on US districting, similar issues arise in other
democracies. For example, the UK is currently reviewing the boundaries for its
parliamentary constituencies. Our work suggests that, even though the UK uses
politically neutral commissions to guide the redistricting process, the results
could disadvantage certain voters due to the cognitive limitations of those
drawing the maps.

Our algorithm is only one possible solution to open and automated districting.
The algorithm selected could be the one that best performs according to an
objective criteria. Different algorithms will provide qualitatively different
geometries, which itself could inform selection. For example, the shortest
splitline algorithm recursively splits a state into districts restricting itself
to North-South and East-West straight lines. The balanced $k$-means algorithm
\citep{chang2014} is very similar to our own algorithm. It minimizes the
standard $k$-means loss function plus an additional weighted term that takes
into account the number of members (i.e., people) in each cluster (i.e.,
district). The range of possible geometries in balanced $k$-means is between
those of the shortest splitline algorithm and our weighted $k$-means. Balanced
$k$-means will create district boundaries that are lines (at any angle, not just
North-South and East-West) to partition the space into a Voronoi diagram. In
contrast, our algorithm, which weights distance by cluster, can form districts
within districts (see \autoref{fig:arizona}) and borders can be curved (see
\autoref{fig:nebraska}). No matter the choice of algorithm, clustering is an
NP-Hard problem such that the optimal solution is not guaranteed unless all
possible assignments are considered \citep{mahajan2012}, which is
computationally impossible in most cases. In practice, random restart with
different initial conditions and other optimization techniques can provide
high-quality solutions.

We believe this automated, yet inclusive and open, approach to redistricting is
preferable to the current system in the USA for which the populace's only remedy
is the court system, which has proven ineffective in this arena. The law and
case history for gerrymandering in the USA is complex and we will not feign to
provide an adequate review here. However, two key points are \textit{a}) courts
are reactive and proceed slowly relative to the pace of election cycles (i.e.,
before any action would be taken, disenfranchisement would have already
occurred); and \textit{b}) the Supreme Court of the United States has never
struck down a politically gerrymandered district \citep{liptak2017}. However,
recently, courts have taken a more active role in addressing cases of
gerrymandering. After centuries of gerrymandering complaints, for the first time
the Supreme Court has agreed to hear a case concerning whether Wisconsin's
partisan gerrymandering is in breach of the First Amendment and the Voting
Rights Act \citep{liptak2017}. Likewise, recent verdicts concerning districting
in North Carolina and Pennsylvania highlight a growing consensus that
politicians should not have a freehand in drawing maps for partisan advantage.

In such legal cases, the concept of voting efficiency, along with comparison to
randomly generated maps \citep{chikina2017}, has prominently
featured \citep{stephanopoulos2015}. The basic concept is that votes for the
losing party in a district are ``wasted" (related to cracking) as well as votes
for the winning party over what is needed to secure victory (related to
packing). Formal measures of efficiency can be readily calculated and
compared \citep{stephanopoulos2015}. Although these measures have their place in
illustrating disparities, we find it preferable to focus on optimising core
principles and values, rather than rarify the status quo and reduce voters to
partisan apparatchiks whose preferences and turnout tendencies are treated as
fixed across election cycles, which they are not.

In contrast to voter efficiency approaches, an algorithm like ours will
naturally lead to cases where groups ``self-gerrymander'', such as when
like-minded communities form in densely populated areas \citep{mccarty2009,
chen2013}. However, it is debatable whether these votes are truly wasted.
Representatives for these relatively homogeneous communities may have a stronger
voice and feel emboldened to advocate for issues that are important to their
community, even when these positions may not be popular on the national stage.
After all, almost by definition, every important social movement, such as the
Civil Rights movement or campaigns for LGBT equality, are not popular at
inception. Nevertheless, concepts like voter efficiency could be included in the
optimisation criteria for algorithms like ours. When faced with complex issues
as to what is fair, the best solution may be the division of labour what we
advocate: humans formalise objective criteria through open discourse and the
computers search for an optimal solution unburdened by human limitations.

\bibliographystyle{ieeetr}

\bibliography{bib}

\begin{thebibliography}{}

\bibitem [\protect \citeauthoryear {%
Altman%
, Amos%
, McDonald%
\BCBL {}\ \BBA {} Smith%
}{%
Altman%
\ \protect \BOthers {.}}{%
{\protect \APACyear {2015}}%
}]{%
altman2015}
\APACinsertmetastar {%
altman2015}%
\begin{APACrefauthors}%
Altman, M.%
, Amos, B.%
, McDonald, M\BPBI P.%
\BCBL {}\ \BBA {} Smith, D\BPBI A.%
\end{APACrefauthors}%
\unskip\
\newblock
\APACrefYearMonthDay{2015}{}{}.
\newblock
{\BBOQ}\APACrefatitle {{Revealing Preferences: Why Gerrymanders are Hard to
  Prove, and What to Do about It}} {{Revealing Preferences: Why Gerrymanders
  are Hard to Prove, and What to Do about It}}.{\BBCQ}
\newblock
\APACjournalVolNumPages{SSRN}{}{}{}.
\newblock
\begin{APACrefDOI} \doi{10.2139/ssrn.2583528} \end{APACrefDOI}
\PrintBackRefs{\CurrentBib}

\bibitem [\protect \citeauthoryear {%
Altman%
, Magar%
, McDonald%
\BCBL {}\ \BBA {} Trelles%
}{%
Altman%
\ \protect \BOthers {.}}{%
{\protect \APACyear {2014}}%
}]{%
altman2014a}
\APACinsertmetastar {%
altman2014a}%
\begin{APACrefauthors}%
Altman, M.%
, Magar, E.%
, McDonald, M\BPBI P.%
\BCBL {}\ \BBA {} Trelles, A.%
\end{APACrefauthors}%
\unskip\
\newblock
\APACrefYearMonthDay{2014}{}{}.
\newblock
{\BBOQ}\APACrefatitle {{The Effects of Automated Redistricting and Partisan
  Strategic Interaction on Representation: The Case of Mexico}} {{The Effects
  of Automated Redistricting and Partisan Strategic Interaction on
  Representation: The Case of Mexico}}.{\BBCQ}
\newblock
\BIn{} \APACrefbtitle {Annual Meeting of the American Political Science
  Association.} {Annual meeting of the american political science association.}
\newblock
\begin{APACrefDOI} \doi{10.2139/ssrn.2486885} \end{APACrefDOI}
\PrintBackRefs{\CurrentBib}

\bibitem [\protect \citeauthoryear {%
Arnold%
}{%
Arnold%
}{%
{\protect \APACyear {2017}}%
}]{%
arnold2017}
\APACinsertmetastar {%
arnold2017}%
\begin{APACrefauthors}%
Arnold, C.%
\end{APACrefauthors}%
\unskip\
\newblock
\APACrefYearMonthDay{2017}{}{}.
\newblock
{\BBOQ}\APACrefatitle {{the Mathematicians Who Want To Save Democracy}} {{the
  Mathematicians Who Want To Save Democracy}}.{\BBCQ}
\newblock
\APACjournalVolNumPages{Nature}{546}{}{8}.
\PrintBackRefs{\CurrentBib}

\bibitem [\protect \citeauthoryear {%
Arthur%
\ \BBA {} Vassilvitskii%
}{%
Arthur%
\ \BBA {} Vassilvitskii%
}{%
{\protect \APACyear {2007}}%
}]{%
arthur2007}
\APACinsertmetastar {%
arthur2007}%
\begin{APACrefauthors}%
Arthur, D.%
\BCBT {}\ \BBA {} Vassilvitskii, S.%
\end{APACrefauthors}%
\unskip\
\newblock
\APACrefYearMonthDay{2007}{}{}.
\newblock
{\BBOQ}\APACrefatitle {{k-means++: The Advantages Of Careful Seeding}}
  {{k-means++: The Advantages Of Careful Seeding}}.{\BBCQ}
\newblock
\BIn{} \APACrefbtitle {{Proceedings of the eighteenth annual ACM-SIAM symposium
  on Discrete algorithms}} {{Proceedings of the eighteenth annual ACM-SIAM
  symposium on Discrete algorithms}}\ (\BPGS\ 1027--1035).
\PrintBackRefs{\CurrentBib}

\bibitem [\protect \citeauthoryear {%
Bazelon%
}{%
Bazelon%
}{%
{\protect \APACyear {2017}}%
}]{%
bazelon2017}
\APACinsertmetastar {%
bazelon2017}%
\begin{APACrefauthors}%
Bazelon, E.%
\end{APACrefauthors}%
\unskip\
\newblock
\APACrefYearMonthDay{2017}{}{}.
\newblock
{\BBOQ}\APACrefatitle {{The New Front in the Gerrymandering Wars: Democracy vs.
  Math}} {{The New Front in the Gerrymandering Wars: Democracy vs.
  Math}}.{\BBCQ}
\newblock
\APACjournalVolNumPages{The New York Times}{}{}{}.
\PrintBackRefs{\CurrentBib}

\bibitem [\protect \citeauthoryear {%
Browdy%
}{%
Browdy%
}{%
{\protect \APACyear {1990}}%
}]{%
browdy1990}
\APACinsertmetastar {%
browdy1990}%
\begin{APACrefauthors}%
Browdy, M\BPBI H.%
\end{APACrefauthors}%
\unskip\
\newblock
\APACrefYearMonthDay{1990}{}{}.
\newblock
{\BBOQ}\APACrefatitle {{Computer Models and Post-Bandemer Redistricting}}
  {{Computer Models and Post-Bandemer Redistricting}}.{\BBCQ}
\newblock
\APACjournalVolNumPages{The Yale Law Journal}{99}{6}{1379--1398}.
\PrintBackRefs{\CurrentBib}

\bibitem [\protect \citeauthoryear {%
{Chang}%
, {Nie}%
, {Ma}%
\BCBL {}\ \BBA {} {Yang}%
}{%
{Chang}%
\ \protect \BOthers {.}}{%
{\protect \APACyear {2014}}%
}]{%
chang2014}
\APACinsertmetastar {%
chang2014}%
\begin{APACrefauthors}%
{Chang}, X.%
, {Nie}, F.%
, {Ma}, Z.%
\BCBL {}\ \BBA {} {Yang}, Y.%
\end{APACrefauthors}%
\unskip\
\newblock
\APACrefYearMonthDay{2014}{{\APACmonth{11}}}{}.
\newblock
{\BBOQ}\APACrefatitle {{Balanced k-Means and Min-Cut Clustering}} {{Balanced
  k-Means and Min-Cut Clustering}}.{\BBCQ}
\newblock
\APACjournalVolNumPages{ArXiv e-prints}{}{}{}.
\PrintBackRefs{\CurrentBib}

\bibitem [\protect \citeauthoryear {%
Chen%
\ \BBA {} Rodden%
}{%
Chen%
\ \BBA {} Rodden%
}{%
{\protect \APACyear {2013}}%
}]{%
chen2013}
\APACinsertmetastar {%
chen2013}%
\begin{APACrefauthors}%
Chen, J.%
\BCBT {}\ \BBA {} Rodden, J.%
\end{APACrefauthors}%
\unskip\
\newblock
\APACrefYearMonthDay{2013}{}{}.
\newblock
{\BBOQ}\APACrefatitle {{Unintentional Gerrymandering: Political Geography And
  Electoral Bias In Legislatures}} {{Unintentional Gerrymandering: Political
  Geography And Electoral Bias In Legislatures}}.{\BBCQ}
\newblock
\APACjournalVolNumPages{Quarterly Journal of Political
  Science}{8}{3}{239--269}.
\PrintBackRefs{\CurrentBib}

\bibitem [\protect \citeauthoryear {%
Chikina%
, Frieze%
\BCBL {}\ \BBA {} Pegden%
}{%
Chikina%
\ \protect \BOthers {.}}{%
{\protect \APACyear {2017}}%
}]{%
chikina2017}
\APACinsertmetastar {%
chikina2017}%
\begin{APACrefauthors}%
Chikina, M.%
, Frieze, A.%
\BCBL {}\ \BBA {} Pegden, W.%
\end{APACrefauthors}%
\unskip\
\newblock
\APACrefYearMonthDay{2017}{}{}.
\newblock
{\BBOQ}\APACrefatitle {{Assessing Significance In A Markov Chain Without
  Mixing}} {{Assessing Significance In A Markov Chain Without Mixing}}.{\BBCQ}
\newblock
\APACjournalVolNumPages{Proceedings of the National Academy of
  Sciences}{114}{11}{2860--2864}.
\newblock
\begin{APACrefURL} \url{http://www.pnas.org/content/114/11/2860}
  \end{APACrefURL}
\newblock
\begin{APACrefDOI} \doi{10.1073/pnas.1617540114} \end{APACrefDOI}
\PrintBackRefs{\CurrentBib}

\bibitem [\protect \citeauthoryear {%
Christensen%
\ \BBA {} Taylor%
}{%
Christensen%
\ \BBA {} Taylor%
}{%
{\protect \APACyear {2001}}%
}]{%
christensen2001}
\APACinsertmetastar {%
christensen2001}%
\begin{APACrefauthors}%
Christensen, M\BPBI E.%
\BCBT {}\ \BBA {} Taylor, M\BPBI G.%
\end{APACrefauthors}%
\unskip\
\newblock
\APACrefYearMonthDay{2001}{}{}.
\newblock
\APACrefbtitle {{Redistricting Committee Report}.} {{Redistricting Committee
  Report}.}
\newblock
\begin{APACrefURL} \url{http://le.utah.gov/interim/2012/pdf/00001272.pdf}
  \end{APACrefURL}
\PrintBackRefs{\CurrentBib}

\bibitem [\protect \citeauthoryear {%
Garc{\'i}a%
\ \protect \BOthers {.}}{%
Garc{\'i}a%
\ \protect \BOthers {.}}{%
{\protect \APACyear {2015}}%
}]{%
garcia2015}
\APACinsertmetastar {%
garcia2015}%
\begin{APACrefauthors}%
Garc{\'i}a, E\BPBI A\BPBI R.%
, Andrade, M\BPBI {\'A}\BPBI G.%
, de-los Cobos-Silva, S\BPBI G.%
, Ponsich, A.%
, Mora-Guti{\'e}rrez, R\BPBI A.%
\BCBL {}\ \BBA {} Lara-Vel{\'a}zquez, P.%
\end{APACrefauthors}%
\unskip\
\newblock
\APACrefYearMonthDay{2015}{}{}.
\newblock
{\BBOQ}\APACrefatitle {{A System for Political Districting in the State of
  Mexico}} {{A System for Political Districting in the State of
  Mexico}}.{\BBCQ}
\newblock
\BIn{} \APACrefbtitle {Advances in Artificial Intelligence and Soft Computing:
  14\textsuperscript{th} Mexican International Conference on Artificial
  Intelligence, MICAI 2015, Cuernavaca, Morelos, Mexico, October 25-31, 2015,
  Proceedings, Part I} {Advances in artificial intelligence and soft computing:
  14\textsuperscript{th} mexican international conference on artificial
  intelligence, micai 2015, cuernavaca, morelos, mexico, october 25-31, 2015,
  proceedings, part i}\ (\BPGS\ 248--259).
\newblock
\APACaddressPublisher{Cham}{Springer International Publishing}.
\PrintBackRefs{\CurrentBib}

\bibitem [\protect \citeauthoryear {%
Graham%
, Joshi%
\BCBL {}\ \BBA {} Pizlo%
}{%
Graham%
\ \protect \BOthers {.}}{%
{\protect \APACyear {2000}}%
}]{%
graham2000}
\APACinsertmetastar {%
graham2000}%
\begin{APACrefauthors}%
Graham, S\BPBI M.%
, Joshi, A.%
\BCBL {}\ \BBA {} Pizlo, Z.%
\end{APACrefauthors}%
\unskip\
\newblock
\APACrefYearMonthDay{2000}{{\APACmonth{10}}}{01}.
\newblock
{\BBOQ}\APACrefatitle {{The Traveling Salesman Problem: A Hierarchical Model}}
  {{The Traveling Salesman Problem: A Hierarchical Model}}.{\BBCQ}
\newblock
\APACjournalVolNumPages{Memory {\&} Cognition}{28}{7}{1191--1204}.
\newblock
\begin{APACrefDOI} \doi{10.3758/BF03211820} \end{APACrefDOI}
\PrintBackRefs{\CurrentBib}

\bibitem [\protect \citeauthoryear {%
Guti{\'{e}}rrez{-}{\'{A}}ndrade%
\ \protect \BOthers {.}}{%
Guti{\'{e}}rrez{-}{\'{A}}ndrade%
\ \protect \BOthers {.}}{%
{\protect \APACyear {2016}}%
}]{%
gutierrezandrade2016}
\APACinsertmetastar {%
gutierrezandrade2016}%
\begin{APACrefauthors}%
Guti{\'{e}}rrez{-}{\'{A}}ndrade, M\BPBI A.%
, Garc{\'{\i}}a, E\BPBI A\BPBI R.%
, de{-}los{-}Cobos{-}Silva, S\BPBI G.%
, Ponsich, A.%
, Guti{\'{e}}rrez, R\BPBI A\BPBI M.%
\BCBL {}\ \BBA {} Vel{\'{a}}zquez, P\BPBI L.%
\end{APACrefauthors}%
\unskip\
\newblock
\APACrefYearMonthDay{2016}{}{}.
\newblock
{\BBOQ}\APACrefatitle {{Redistricting in Mexico}} {{Redistricting in
  Mexico}}.{\BBCQ}
\newblock
\BIn{} A.~Fink, A.~F{\"{u}}genschuh\BCBL {}\ \BBA {} M\BPBI J.~Geiger\ (\BEDS),
  \APACrefbtitle {Operations Research Proceedings 2016, Selected Papers of the
  Annual International Conference of the German Operations Research Society
  (GOR), Helmut Schmidt University Hamburg, Germany, August 30 - September 2,
  2016.} {Operations research proceedings 2016, selected papers of the annual
  international conference of the german operations research society (gor),
  helmut schmidt university hamburg, germany, august 30 - september 2, 2016.}\
  (\BPGS\ 301--306).
\newblock
\APACaddressPublisher{}{Springer}.
\newblock
\begin{APACrefDOI} \doi{10.1007/978-3-319-55702-1_40} \end{APACrefDOI}
\PrintBackRefs{\CurrentBib}

\bibitem [\protect \citeauthoryear {%
Hess%
, Weaver%
, Siegfeldt%
, Whelan%
\BCBL {}\ \BBA {} Zitlau%
}{%
Hess%
\ \protect \BOthers {.}}{%
{\protect \APACyear {1965}}%
}]{%
hess1965}
\APACinsertmetastar {%
hess1965}%
\begin{APACrefauthors}%
Hess, S\BPBI W.%
, Weaver, J\BPBI B.%
, Siegfeldt, H\BPBI J.%
, Whelan, J\BPBI N.%
\BCBL {}\ \BBA {} Zitlau, P\BPBI A.%
\end{APACrefauthors}%
\unskip\
\newblock
\APACrefYearMonthDay{1965}{}{}.
\newblock
{\BBOQ}\APACrefatitle {{Nonpartisan Political Redistricting by Computer}}
  {{Nonpartisan Political Redistricting by Computer}}.{\BBCQ}
\newblock
\APACjournalVolNumPages{Operations Research}{13}{6}{998--1006}.
\newblock
\begin{APACrefDOI} \doi{10.1287/opre.13.6.998} \end{APACrefDOI}
\PrintBackRefs{\CurrentBib}

\bibitem [\protect \citeauthoryear {%
Issacharoff%
}{%
Issacharoff%
}{%
{\protect \APACyear {2002}}%
}]{%
issacharoff2002}
\APACinsertmetastar {%
issacharoff2002}%
\begin{APACrefauthors}%
Issacharoff, S.%
\end{APACrefauthors}%
\unskip\
\newblock
\APACrefYearMonthDay{2002}{}{}.
\newblock
{\BBOQ}\APACrefatitle {{Gerrymandering and Political Cartels}} {{Gerrymandering
  and Political Cartels}}.{\BBCQ}
\newblock
\APACjournalVolNumPages{Harvard Law Review}{116}{2}{593--648}.
\PrintBackRefs{\CurrentBib}

\bibitem [\protect \citeauthoryear {%
Levitt%
}{%
Levitt%
}{%
{\protect \APACyear {2011}}%
}]{%
levitt2011}
\APACinsertmetastar {%
levitt2011}%
\begin{APACrefauthors}%
Levitt, J.%
\end{APACrefauthors}%
\unskip\
\newblock
\APACrefYearMonthDay{2011}{}{}.
\newblock
{\BBOQ}\APACrefatitle {{The Legal Context for Scientific Redistricting
  Analysis}} {{The Legal Context for Scientific Redistricting
  Analysis}}.{\BBCQ}
\newblock
\APACjournalVolNumPages{SSRN}{}{}{}.
\PrintBackRefs{\CurrentBib}

\bibitem [\protect \citeauthoryear {%
Liptak%
}{%
Liptak%
}{%
{\protect \APACyear {2017}}%
}]{%
liptak2017}
\APACinsertmetastar {%
liptak2017}%
\begin{APACrefauthors}%
Liptak, A.%
\end{APACrefauthors}%
\unskip\
\newblock
\APACrefYearMonthDay{2017}{}{}.
\newblock
{\BBOQ}\APACrefatitle {{Justices to Hear Major Challenge to Partisan
  Gerrymandering}} {{Justices to Hear Major Challenge to Partisan
  Gerrymandering}}.{\BBCQ}
\newblock
\APACjournalVolNumPages{The New York Times}{}{}{}.
\PrintBackRefs{\CurrentBib}

\bibitem [\protect \citeauthoryear {%
Lloyd%
}{%
Lloyd%
}{%
{\protect \APACyear {1982}}%
}]{%
lloyd1982}
\APACinsertmetastar {%
lloyd1982}%
\begin{APACrefauthors}%
Lloyd, S.%
\end{APACrefauthors}%
\unskip\
\newblock
\APACrefYearMonthDay{1982}{}{}.
\newblock
{\BBOQ}\APACrefatitle {{Least Squares Quantization in PCM}} {{Least Squares
  Quantization in PCM}}.{\BBCQ}
\newblock
\APACjournalVolNumPages{IEEE transactions on information
  theory}{28}{2}{129--137}.
\PrintBackRefs{\CurrentBib}

\bibitem [\protect \citeauthoryear {%
Mahajan%
, Nimbhorkar%
\BCBL {}\ \BBA {} Varadarajan%
}{%
Mahajan%
\ \protect \BOthers {.}}{%
{\protect \APACyear {2012}}%
}]{%
mahajan2012}
\APACinsertmetastar {%
mahajan2012}%
\begin{APACrefauthors}%
Mahajan, M.%
, Nimbhorkar, P.%
\BCBL {}\ \BBA {} Varadarajan, K.%
\end{APACrefauthors}%
\unskip\
\newblock
\APACrefYearMonthDay{2012}{}{}.
\newblock
{\BBOQ}\APACrefatitle {{The Planar k-means Problem is NP-hard}} {{The Planar
  k-means Problem is NP-hard}}.{\BBCQ}
\newblock
\APACjournalVolNumPages{Theoretical Computer Science}{442}{}{13--21}.
\newblock
\begin{APACrefURL}
  \url{http://www.sciencedirect.com/science/article/pii/S0304397510003269}
  \end{APACrefURL}
\newblock
\APACrefnote{Special Issue on the Workshop on Algorithms and Computation
  (WALCOM 2009)}
\newblock
\begin{APACrefDOI} \doi{10.1016/j.tcs.2010.05.034} \end{APACrefDOI}
\PrintBackRefs{\CurrentBib}

\bibitem [\protect \citeauthoryear {%
Martis%
}{%
Martis%
}{%
{\protect \APACyear {2008}}%
}]{%
Martis2008}
\APACinsertmetastar {%
Martis2008}%
\begin{APACrefauthors}%
Martis, K\BPBI C.%
\end{APACrefauthors}%
\unskip\
\newblock
\APACrefYearMonthDay{2008}{}{}.
\newblock
{\BBOQ}\APACrefatitle {The original gerrymander} {The original
  gerrymander}.{\BBCQ}
\newblock
\APACjournalVolNumPages{Political Geography}{27}{8}{833 - 839}.
\newblock
\begin{APACrefURL}
  \url{http://www.sciencedirect.com/science/article/pii/S0962629808000954}
  \end{APACrefURL}
\newblock
\begin{APACrefDOI} \doi{https://doi.org/10.1016/j.polgeo.2008.09.003}
  \end{APACrefDOI}
\PrintBackRefs{\CurrentBib}

\bibitem [\protect \citeauthoryear {%
McCarty%
, Poole%
\BCBL {}\ \BBA {} Rosenthal%
}{%
McCarty%
\ \protect \BOthers {.}}{%
{\protect \APACyear {2009}}%
}]{%
mccarty2009}
\APACinsertmetastar {%
mccarty2009}%
\begin{APACrefauthors}%
McCarty, N.%
, Poole, K\BPBI T.%
\BCBL {}\ \BBA {} Rosenthal, H.%
\end{APACrefauthors}%
\unskip\
\newblock
\APACrefYearMonthDay{2009}{}{}.
\newblock
{\BBOQ}\APACrefatitle {{Does Gerrymandering Cause Polarization?}} {{Does
  Gerrymandering Cause Polarization?}}{\BBCQ}
\newblock
\APACjournalVolNumPages{American Journal of Political
  Science}{53}{3}{666--680}.
\PrintBackRefs{\CurrentBib}

\bibitem [\protect \citeauthoryear {%
Myers%
}{%
Myers%
}{%
{\protect \APACyear {2002}}%
}]{%
myers2002}
\APACinsertmetastar {%
myers2002}%
\begin{APACrefauthors}%
Myers, D\BPBI G.%
\end{APACrefauthors}%
\unskip\
\newblock
\APACrefYear{2002}.
\newblock
\APACrefbtitle {{Intuition: Its Powers and Perils}} {{Intuition: Its Powers and
  Perils}}.
\newblock
\APACaddressPublisher{}{Yale University Press}.
\PrintBackRefs{\CurrentBib}

\bibitem [\protect \citeauthoryear {%
Palmer%
}{%
Palmer%
}{%
{\protect \APACyear {1990}}%
}]{%
Palmer1990}
\APACinsertmetastar {%
Palmer1990}%
\begin{APACrefauthors}%
Palmer, S\BPBI E.%
\end{APACrefauthors}%
\unskip\
\newblock
\APACrefYearMonthDay{1990}{}{}.
\newblock
{\BBOQ}\APACrefatitle {{Modern Theories of Gestalt Perception}} {{Modern
  Theories of Gestalt Perception}}.{\BBCQ}
\newblock
\APACjournalVolNumPages{Mind and Language}{5}{4}{289--323}.
\PrintBackRefs{\CurrentBib}

\bibitem [\protect \citeauthoryear {%
Ponsich%
\ \protect \BOthers {.}}{%
Ponsich%
\ \protect \BOthers {.}}{%
{\protect \APACyear {2017}}%
}]{%
ponsich2017}
\APACinsertmetastar {%
ponsich2017}%
\begin{APACrefauthors}%
Ponsich, A.%
, Garc\'{\i}a, E\BPBI A\BPBI R.%
, Guti{\'e}rrez, R\BPBI A\BPBI M.%
, Silva, S\BPBI G\BPBI d\BHBI l\BHBI C.%
, Andrade, M\BPBI A\BPBI G.%
\BCBL {}\ \BBA {} Vel\'{a}zquez, P\BPBI L.%
\end{APACrefauthors}%
\unskip\
\newblock
\APACrefYearMonthDay{2017}{}{}.
\newblock
{\BBOQ}\APACrefatitle {{Solving Electoral Zone Design Problems with NSGA-II:
  Application to Redistricting in Mexico}} {{Solving Electoral Zone Design
  Problems with NSGA-II: Application to Redistricting in Mexico}}.{\BBCQ}
\newblock
\BIn{} \APACrefbtitle {Proceedings of the Genetic and Evolutionary Computation
  Conference Companion} {Proceedings of the genetic and evolutionary
  computation conference companion}\ (\BPGS\ 159--160).
\newblock
\APACaddressPublisher{New York, NY, USA}{ACM}.
\newblock
\begin{APACrefDOI} \doi{10.1145/3067695.3076103} \end{APACrefDOI}
\PrintBackRefs{\CurrentBib}

\bibitem [\protect \citeauthoryear {%
Stephanopoulos%
\ \BBA {} McGhee%
}{%
Stephanopoulos%
\ \BBA {} McGhee%
}{%
{\protect \APACyear {2015}}%
}]{%
stephanopoulos2015}
\APACinsertmetastar {%
stephanopoulos2015}%
\begin{APACrefauthors}%
Stephanopoulos, N\BPBI O.%
\BCBT {}\ \BBA {} McGhee, E\BPBI M.%
\end{APACrefauthors}%
\unskip\
\newblock
\APACrefYearMonthDay{2015}{}{}.
\newblock
{\BBOQ}\APACrefatitle {{Partisan Gerrymandering And The Efficiency Gap}}
  {{Partisan Gerrymandering And The Efficiency Gap}}.{\BBCQ}
\newblock
\APACjournalVolNumPages{The University of Chicago Law Review}{}{}{831--900}.
\PrintBackRefs{\CurrentBib}

\bibitem [\protect \citeauthoryear {%
Taylor%
}{%
Taylor%
}{%
{\protect \APACyear {2016}}%
}]{%
taylor2016}
\APACinsertmetastar {%
taylor2016}%
\begin{APACrefauthors}%
Taylor, D.%
\end{APACrefauthors}%
\unskip\
\newblock
\APACrefYearMonthDay{2016}{}{}.
\newblock
{\BBOQ}\APACrefatitle {{Gerrymandering, Explained}} {{Gerrymandering,
  Explained}}.{\BBCQ}
\newblock
\APACjournalVolNumPages{The Washington Post}{}{}{}.
\newblock
\begin{APACrefURL}
  \url{https://www.washingtonpost.com/video/business/gerrymandering-explained/2016/04/21/e447f5c2-07fe-11e6-bfed-ef65dff5970d_video.html}
  \end{APACrefURL}
\PrintBackRefs{\CurrentBib}

\bibitem [\protect \citeauthoryear {%
Vickrey%
}{%
Vickrey%
}{%
{\protect \APACyear {1961}}%
}]{%
vickrey1961}
\APACinsertmetastar {%
vickrey1961}%
\begin{APACrefauthors}%
Vickrey, W.%
\end{APACrefauthors}%
\unskip\
\newblock
\APACrefYearMonthDay{1961}{}{}.
\newblock
{\BBOQ}\APACrefatitle {{On the Prevention of Gerrymandering}} {{On the
  Prevention of Gerrymandering}}.{\BBCQ}
\newblock
\APACjournalVolNumPages{Political Science Quarterly}{}{}{}.
\PrintBackRefs{\CurrentBib}

\end{thebibliography}

\end{document}